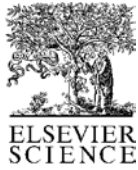
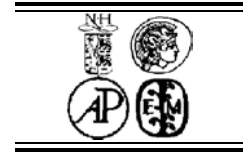

# Thermal properties of MgB$_2$: the effect of disorder on gap amplitudes and relaxation times of π and σ bands

M. Putti,[a] E. Galleani,[a] I. Pallecchi,[a] C. Bernini,[a] P. Manfrinetti,[b] A. Palenzona,[b] M. Affronte,[c]

[a]*INFM-LAMIA/CNR, Dipartimento di Fisica, Via Dodecaneso 33, 16146 Genova, Italy*

[b]*INFM-LAMIA, Dipartimento di Chimica e Chimica Industriale, Via Dodecaneso 31, 16146 Genova, Italy*

[c]*INFM-NRC S$^3$, Dipartimento di Fisica, via G.Campi 213/A, I-41100 Modena, Italy*

**Abstract**

We present thermal conductivity and specific heat measurements on MgB$_2$ and Mg-AlB$_2$ samples. Thermal properties have been analysed by using a two-gap model in order to estimate the gap amplitudes, $\Delta(0)_\pi$ and $\Delta(0)_\sigma$ and the intra-band scattering rates, $\Gamma_{\sigma\sigma}$ and $\Gamma_{\pi\pi}$. As a function of Al doping and disorder $\Delta(0)_\sigma$ rapidly decreases, while $\Delta(0)_\pi$ is rather constant. $\Gamma_{\sigma\sigma}$ and $\Gamma_{\pi\pi}$ are increased by the disorder, being $\Gamma_{\pi\pi}$ more affected than $\Gamma_{\sigma\sigma}$.

Despite its conventional nature, superconductivity in MgB2 shows several unusual features that can be ascribed to the presence of two gaps with different amplitudes [1]: the larger gap, $\Delta(0)_\sigma$, is associated with two-dimensional σ bands and the smaller one, $\Delta(0)_\pi$, is associated with three-dimensional π bands. Recently, it has been evidenced that inter-band scattering by non-magnetic impurities in a multigap superconductor [2] has a pair-breaking effect and suppresses T$_c$ as scattering by magnetic impurities does in a regular superconductor. In MgB$_2$ inter-band impurity scattering is strongly reduced due to the different symmetry of σ and π bands, therefore σ and π bands behave as conduction channels in parallel [3]. Diffusion from one band to the other is inhibited not only to quasiparticles, but to Cooper pairs as well and this preserves the T$_c$ suppression and the large anisotropy of the energy gaps.

In this paper some results obtained by thermal conductivity [4] and specific heat [5] are summarized. In particular we analyze the evolution of the gap amplitudes, intra-band ($\Gamma_{\sigma\sigma}$, $\Gamma_{\pi\pi}$) and inter-band ($\Gamma_{\sigma\pi}$) impurity scattering rates as a function of Al doping and disorder.

The thermal properties in the superconducting state has been analyzed within a two-gap model. In particular the specific heat is given by [6]:

$$C_{es}(T)/C_{en}(T) = xc(t,\alpha_\pi) + (1-x)c(t,\alpha_\sigma) \quad (1)$$

where $C_{en}(T)=\gamma T$ (γ is the Sommerfeld constant), $x=\gamma_\pi/\gamma$ and $(1-x)=\gamma_\sigma/\gamma$ are the energy fraction condensing in each bands, $c$ is the normalized BCS specific heat function, $t=T/T_c$ is the reduced temperature and $\alpha_\pi=\Delta(0)_\pi/k_B T_c$ and $\alpha_\sigma=\Delta(0)_\sigma/k_B T_c$ are the reduced gap. A similar equation can be written for the thermal conductivity [4]:

$$K_{es}(T)/K_{en}(T) = yg(t,\alpha_\pi) + (1-y)g(t,\alpha_\sigma) \quad (2)$$

where the relative weights $y$ and $(1-y)$ in this case represent the energy fractions carried by the π and σ bands, respectively and $g$ is the normalized BCS thermal conductivity function.

Specific heat measurements were performed on high quality polycrystalline Mg$_{1-f}$Al$_f$B$_2$ samples with f=0, 0.1,

0.2, 0.3 [5]. Thermal conductivity measurements were performed on bulk samples obtained by a one step technique [4]. Three specimens were prepared for transport measurements: one from crystalline boron (MGB), one by using enriched crystalline $^{11}$B (MGB11), one with 5% Al in the site of Mg (MGALB).

Fig. 1 shows the gap amplitudes as a function of Al content as estimated by specific heat measurements. For all the samples we estimated $x\sim0.5$ as expected by band structure calculation. $\Delta(0)_\sigma$ linearly decreases, while $\Delta(0)_\pi$ remains constant up to f=0.2. Therefore, $\Delta(0)_\sigma$ and $\Delta(0)_\pi$ which differ more than a factor three in pure MgB$_2$, tend to the same value at f~0.3. In fig. 1 we report also a theoretical evaluation of the gap amplitude as a function of the Al doping [7] where the band filling effect is considered. The poor agreement between the theoretical prediction and the experimental data suggests that the changes in the electronic structure do not completely account for the changes in the energy gaps. We believe that also the disorder introduced by the Al doping, increasing the inter-band scattering rate, plays an important role for the degradation of the superconducting properties. The evolution of the two gaps as $\Gamma_{\sigma\pi}$ increases [8] can be compared with our data. As $\Gamma_{\sigma\pi}$ increases the larger gap $\Delta(0)_\sigma$ decreases, while the small gap $\Delta(0)_\pi$ grows as consequence of the fact that a larger number of pairs is scattered in the second band. Eventually the strong scattering limit is achieved when $\Gamma_{\sigma\pi}$ becomes larger than the relevant phonon frequency (~75 meV) and the two gaps merge into one.

To estimate the impurity scattering rates we analyze the thermal conductivity of the samples MGB11, MGB and MGALB. By the equation (2), $\Gamma_{\sigma\sigma}$ and $\Gamma_{\pi\pi}$, which are inversely proportional to the weights y and *(1-y)*, can be calculated. In fig 2 (a), $\Gamma_{\sigma\sigma}$ and $\Gamma_{\pi\pi}$ are plotted as a function of the residual resistivity $\rho_0$, in order to emphasize the dependence of these parameters on the disorder. It can be seen that both the intra-band scattering rates increase with the disorder, but $\Gamma_{\pi\pi}$, with larger sensitivity, and becomes two times larger than $\Gamma_{\sigma\sigma}$ in MGALB. This result is very reasonable; in fact Al substitution in the Mg site limits the mobility of $\pi$ carriers, but not that of $\sigma$ carriers which move in the B planes. Rough estimation of inter-band scattering rate $\Gamma_{\sigma\pi}$ can be obtained by the relation $(\delta T_c/T_c) \approx -(\Gamma_{\sigma\pi}/k_B T_c)(\delta\Delta^2/\overline{\Delta^2})$ [2], where $\delta T_c$ is the $T_c$ reduction in respect to the optimal value, $\delta\Delta=\Delta_\sigma - \Delta_\pi$ and $\overline{\Delta^2} = \Delta_\sigma^2 + \Delta_\pi^2$. For MGB11 and MGB, which have $T_c$ =38.7 K and 38.9 K, close to the optimal value an upper limit could be $\Gamma_{\sigma\pi}<0.1$ meV. The upper limit for MGALB ($T_c$=35 K) is obtained assuming that the $T_c$ reduction is due only to inter-band impurity scattering (and not to band filling). In such a way we obtain $\Gamma_{\sigma\pi} < 2$ meV. Thus for all the samples we find $\Gamma_{\sigma\pi}<<\Gamma_{\sigma\sigma}, \Gamma_{\pi\pi}$ as expected due to the different parity of $\sigma$ and $\pi$ bands.

Finally in fig. 2 (b) we plot $\Delta(0)_\sigma$ and $\Delta(0)_\pi$ as a function of $\rho_0$. The gap behavior with the disorder, in the limited region we have investigated, looks similar to the behavior with the Al doping (fig.1). Actually, for a more complete analysis it is important to collect data on sample with high level of disorder.

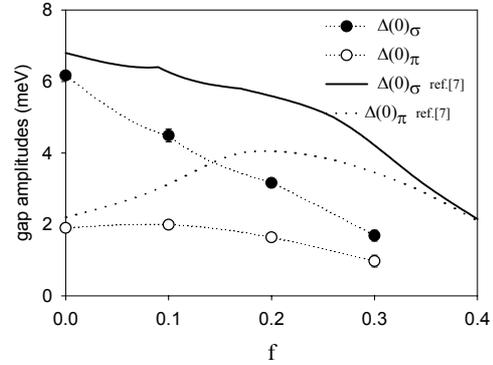

Fig.1 $\Delta(0)_\sigma$ and $\Delta(0)_\pi$ as a function of Al doping

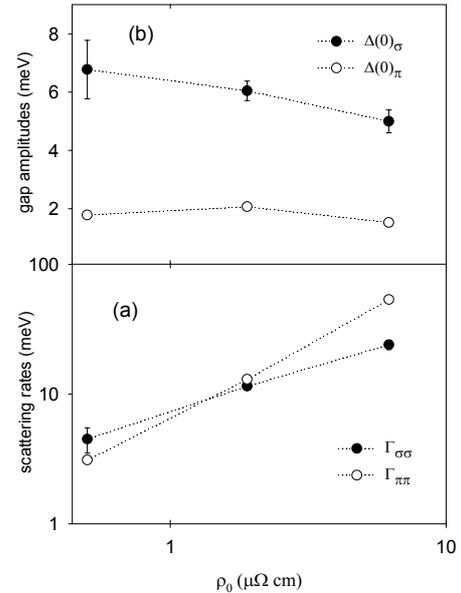

Fig.2 (a) $\Gamma_{\sigma\sigma}$ and $\Gamma_{\pi\pi}$ and (b) $\Delta(0)_\sigma$ and $\Delta(0)_\pi$ as a function of $\rho_0$; the samples are MGB11($\rho_0\sim0.5\mu\Omega$cm), MGB ($\rho_0\sim2\mu\Omega$cm) and MGALB ($\rho_0\sim6\mu\Omega$cm).